\documentclass[aps,prl,reprint,floatfix]{revtex4-2}

\usepackage{graphicx}
\usepackage{textgreek}
\usepackage{siunitx}
\usepackage{xcolor}

\usepackage[version=3]{mhchem} % Formula subscripts using \ce{}

\begin{document}
\author{Nicolas Rendler}
\author{Audrey Scognamiglio}
\affiliation{Institute of Physics, University of Freiburg, Hermann-Herder-Str. 3, 79104 Freiburg, Germany}
\author{Manuel Barranco}
\affiliation{Departament FQA, Facultat de F\'{i}sica, Universitat de Barcelona, Diagonal 645, 08028 Barcelona, Spain}
\affiliation{Institute of Nanoscience and Nanotechnology (IN2UB), Universitat de Barcelona, Barcelona, Spain}
\author{Marti P\'{i}}
\affiliation{Departament FQA, Facultat de F\'{i}sica, Universitat de Barcelona, Diagonal 645, 08028 Barcelona, Spain}
\affiliation{Institute of Nanoscience and Nanotechnology (IN2UB), Universitat de Barcelona, Barcelona, Spain}
\author{Nadine Halberstadt}
\affiliation{Laboratoire des Collisions, Agr\'{e}gats, R\'{e}activit\'{e}, IRSAMC, Universit\'{e} de Toulouse, CNRS UMR 5589, 31062 Toulouse Cedex 09, France}
\author{Katrin Dulitz}
\author{Frank Stienkemeier}
\affiliation{Institute of Physics, University of Freiburg, Hermann-Herder-Str. 3, 79104 Freiburg, Germany}
\email{stienkemeier@uni-freiburg.de}

%%%%%%%%%%%%%%%%%%%%%%%%%%%%%%%%%%%%%%%%%%%%%%%%%%%%%%%%%%%%%%%%%%%%%
%% The document title should be given as usual. Some journals require
%% a running title from the author: this should be supplied as an
%% optional argument to \title.
%%%%%%%%%%%%%%%%%%%%%%%%%%%%%%%%%%%%%%%%%%%%%%%%%%%%%%%%%%%%%%%%%%%%%
\title{Dynamics of Photoexcited Cs Atoms Attached to Helium Nanodroplets}

%%%%%%%%%%%%%%%%%%%%%%%%%%%%%%%%%%%%%%%%%%%%%%%%%%%%%%%%%%%%%%%%%%%%%
%% Some journals require a list of abbreviations or keywords to be
%% supplied. These should be set up here, and will be printed after
%% the title and author information, if needed.
%%%%%%%%%%%%%%%%%%%%%%%%%%%%%%%%%%%%%%%%%%%%%%%%%%%%%%%%%%%%%%%%%%%%%


%%%%%%%%%%%%%%%%%%%%%%%%%%%%%%%%%%%%%%%%%%%%%%%%%%%%%%%%%%%%%%%%%%%%%
%% The manuscript does not need to include \maketitle, which is
%% executed automatically.
%%%%%%%%%%%%%%%%%%%%%%%%%%%%%%%%%%%%%%%%%%%%%%%%%%%%%%%%%%%%%%%%%%%%%

\begin{abstract}
We present an experimental study of the dynamics following the photoexcitation and subsequent photoionization of single Cs atoms on the surface of helium nanodroplets. The dynamics of excited Cs atom desorption and readsorption as well as CsHe exciplex formation are measured using femtosecond pump-probe velocity map imaging spectroscopy and ion-time-of-flight spectrometry. The time scales for the desorption of excited Cs atoms off helium nanodroplets as well as the time scales for CsHe exciplex formation are experimentally determined for the 6p states of Cs. For the 6p $^2\Pi_ {1/2}$ state, our results confirm that the excited Cs atoms only desorb from the nanodroplet when the excitation wavenumber is blue-shifted from the $6p\,^2\Pi_ {1/2} \leftarrow 6s\,^2\Sigma_ {1/2}$ resonance. 
Our results suggest that the dynamics following excitation to the 6p $^2\Pi_ {3/2}$ state can be described by an evaporation-like desorption mechanism, whereas the dynamics arising from excitation to the 6p $^2\Sigma_ {1/2}$ state is indicative for a more impulsive desorption process. Furthermore, our results suggest a helium-induced spin-orbit relaxation from the 6p $^2\Sigma_ {1/2}$ state to the 6p $^2\Pi_ {1/2}$ state. Our findings largely agree with the results of time-dependent $^4$He-density-functional theory (DFT) simulations published earlier [\textit{Eur. Phys. J. D} \textbf{2019}, \textit{73}, 94].
\end{abstract}
\maketitle

\section{Introduction}
\label{intro}
Due to the inertness and the exceptional characteristics of superfluid helium (He), He nanodroplets are widely used as a weakly perturbing matrices for atoms, molecules, aggregates and clusters. The applications of He nanodroplets to experimental investigations are extensive and include, for instance, high-resolution frequency-resolved spectroscopy\,\cite{Toennies.2004, Stienkemeier.2001, Choi.2006}, time-resolved spectroscopy\,\cite{Stienkemeier.2006, Mudrich.2014, Ziemkiewicz.2015}, chemical reaction\,\cite{Mauracher.2018} and nanoplasma formation studies\,\cite{Krishnan.2011, Mudrich.2014, Heidenreich.2016}, and the study of nanostructure formation and deposition\,\cite{Mozhayskiy.2007, Ernst.2021, Emery.2013}. Despite the inertness of the He nanodroplets, dopant-host dynamics can be triggered by the electronic excitation and subsequent ionization of dopants which can involve, for example, the desorption of a dopant from the droplet, the broadening and shift of spectral features, the formation of exciplexes and the ignition of a nanoplasma\,\cite{Heidenreich.2016}. This motivates further studies aimed at understanding these dynamics in detail. 
The dynamics of photoexcited alkali metal atom dopants on the surface of He nanodroplets have already been intensively investigated in both theoretical and experimental studies. For the theoretical studies, a pseudo-diatomic model was used, in which the He nanodroplet, He$_N$ (where $N$ is the number of He atoms contained in the droplet), is treated like a heavy constituent atom of the alkali atom-He$_N$ complex. This model, in which the potential energy curves of the alkali atom are calculated as a function of the atom's distance $R$ from the center of the He nanodroplet, has proven to be suitable for predicting the dynamics of such systems to a good extent\,\cite{Vangerow.2015, Vangerow.2017, Stienkemeier.1996, Bunermann.2007}.   
Due to the weak attraction between the alkali metal atom and the He atoms of the nanodroplet, alkali atoms reside in a dimple at the surface of the droplet\,\cite{Ancilotto.1995, Dalfovo.1994}. After electronic excitation of the alkali atom, Pauli repulsion between the electronic distributions of the excited alkali atom and the surrounding ground-state He atoms usually leads to the desorption of the excited alkali atom, with exceptions for some of the lowest excited states of Rb and Cs\,\cite{Theisen.2011, Theisen.2011c, Aubock.2008}.  
The desorption and solvation dynamics of excited alkali atoms from He nanodroplets depend on the time delay between the excitation and ionization steps, on the specific excited state and on the excitation wavenumber. The involved time scales can thus vary significantly in between the different alkali-atom species. Therefore, separate studies for all alkali atom He-nanodroplet systems are needed to understand the underlying dynamics.

As a response to photoexcitation, alkali-He exciplexes can be formed on the surface of He nanodroplets due to local minima in the pseudo-diatomic potentials of the dopant-droplet system\,\cite{Reho.1997}. Exciplex formation has already been observed for Li\,\cite{Enomoto.2004}, Na\,\cite{Reho.1997}, K\,\cite{Schulz.2001, Enomoto.2004, Reho.2000, Reho.2000b, Reho.1997}, Rb\,\cite{Coppens.2018}, Cs\,\cite{Nettels.2005, Moroshkin.2006},  Cr\,\cite{Koch.2014} and Ag\,\cite{Persson.1996}. Due to the higher number of degrees of freedom, the dynamics of exciplexes attached to He nanodroplets are more complicated as compared to atom-droplet dynamics. Despite the inertness of the superfluid He, it is well known that the He environment can induce shifts and strong broadenings of atomic transitions. Transitions that are electric-dipole forbidden in free atoms can become allowed in the droplet-bound atom due to the reduced symmetry of the dopant-He-nanodroplet system. In previous experiments of Cs-doped He nanodroplets, the desorption and solvation dynamics of photoexcited Cs atoms have been investigated using laser-induced fluorescence spectroscopy, beam depletion and time-of-flight mass spectrometry\,\cite{Theisen.2011}. The experimental results presented here were obtained using femtosecond-pump-probe spectroscopy and thus provide additional insights into the time scales of the involved dynamics.   

\begin{figure}
\includegraphics[width=0.8\columnwidth]{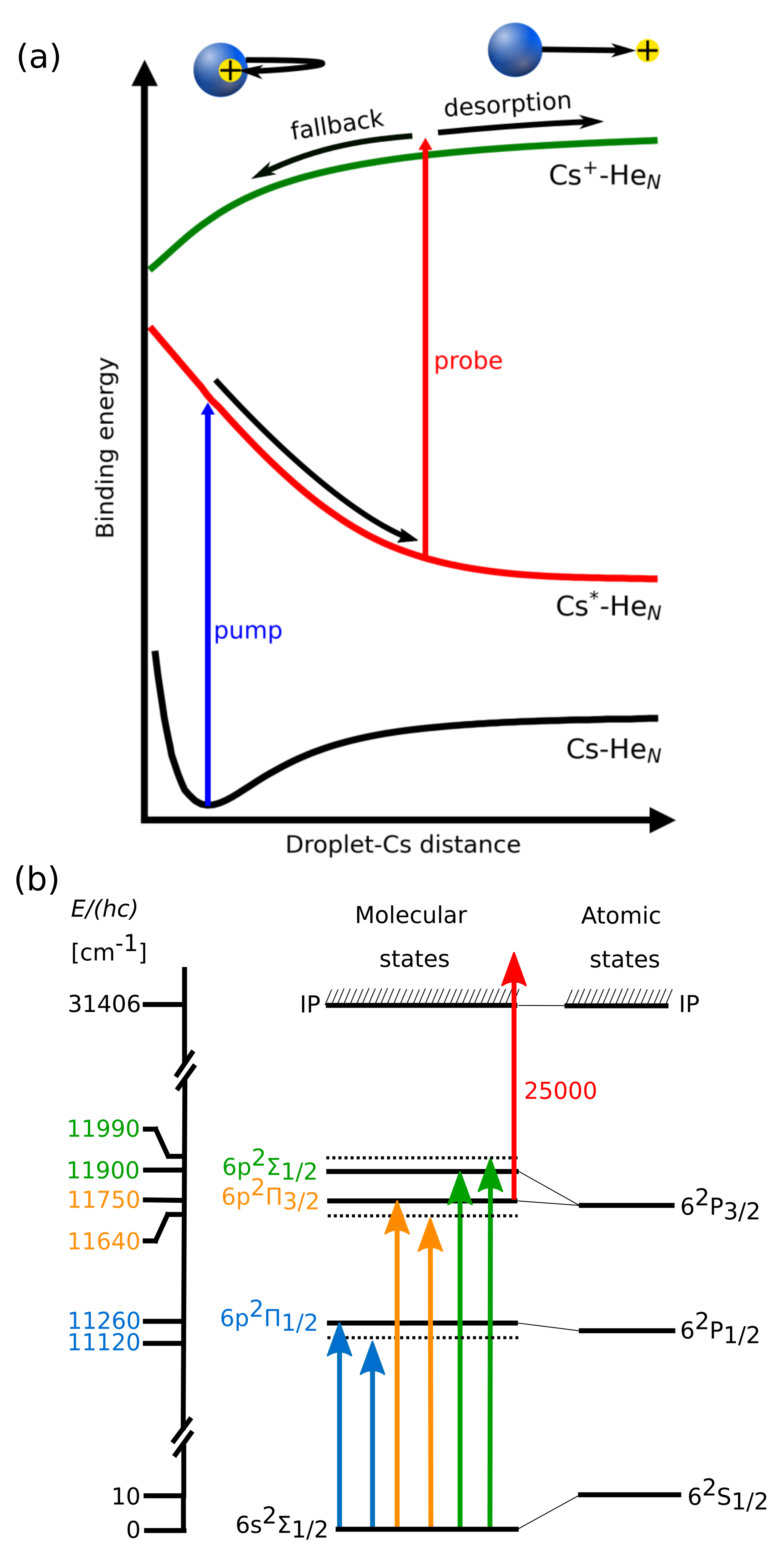}
\caption{(a) Schematic pseudo-diatomic potential energy diagram of a Cs-He$_N$ nanodroplet complex and illustration of the femtosecond pump-probe photoionization scheme used in the experiments.
(b) Correlation diagram showing the lowest atomic states of Cs (right) and the corresponding molecular states of a Cs-He$_N$ complex in the approximation of a pseudo-diatomic model (left). The pump laser wavenumbers $E_{\mathrm{pump}}/(hc)$ used to excite the 6p $^{2}\Pi _{1/2}$, 6p $^{2}\Pi _{3/2}$, and 6p $^{2}\Sigma _{1/2}$ states are indicated by blue, orange, and green arrows, respectively. The probe-laser wavenumber $E_{\mathrm{probe}}/(hc)$ (red color) is depicted for the ionization of the 6p $^{2}\Pi _{3/2}$ state only. The wavenumber axis of the diagram is not to scale.}
\label{fig:fallback schematic+levels scheme}
\end{figure}
The interaction between an electronically excited Cs atom and a He nanodroplet is often repulsive because of the high delocalization of the valence electron in excited-state Cs\,\cite{Callegari.2011, Reho.2000b}. The 6p $^{2}\Pi _{1/2}$, 6p $^{2}\Pi _{3/2}$ and 6p $^{2}\Sigma _{1/2}$ molecular states, which are excited near the 6p $\leftarrow$ 6s transitions in atomic Cs, have bound-state character, but resonant excitation near the equilibrium position of the Cs-He$_N$ complex is predicted to produce excited-state Cs atoms in molecular states far above the dissociation limit, so that a desorption of the excited Cs atom from the droplet is expected\,\cite{Coppens.2019}. 
Using time-dependent $^4$He-DFT simulations, the same authors have found that excited-state Cs atoms desorb after excitation to the 6p $^{2}\Sigma _{1/2}$ state, while for the 6p $^{2}\Pi_{1/2}$ state, the atoms are predicted to desorb from the droplet when excitation is close to the maximum of the $^{2}\Pi_{1/2}$ band, at $E_{\mathrm{pump}}/(hc)$ = 11260\,cm$^{-1}$, and remain attached when excitation is done at lower wavenumbers.
The results of the theoretical studies also imply that CsHe exciplex formation is supported for excitation to the 6p $^{2}\Pi _{3/2}$ and 6p $^{2}\Pi _{1/2}$ states only, and that the desorption of the exciplex in the 6p $^{2}\Pi _{3/2}$ state is due to nonradiative relaxation to the 6p $^{2}\Pi _{1/2}$ state\,\cite{Coppens.2019}.

Figure \ref{fig:fallback schematic+levels scheme}a shows a schematic illustration of the dynamics following the photoexcitation and subsequent photo-ionization of a Cs atom on the surface of a He nanodroplet. Schematic pseudo-diatomic potential curves are shown as a function of distance between the He nanodroplet and the Cs atom. The excitation (pump) step, indicated by the blue arrow, leads to the excitation of a wavepacket in the excited molecular state. In the case depicted here, the excited state is repulsive, so that the excited-state Cs atom desorbs from the droplet. In the ionization (probe) step (red arrow), a Cs$^+$-He$_N$ molecular ion is formed. This potential is attractive because of the attractive force between the positive charge of the Cs$^+$ ion and the He nanodroplet\,\cite{Toennies.1998}. During the time delay between the excitation and ionization steps, the excited wavepacket evolves along the excited state of the potential energy curve. Depending on this time delay, two possible scenarios can occur. The first possibility is that the excited Cs atom has gained enough kinetic energy and has reached a distance at which the droplet attraction is weaker, so that the ion formed after photo-ionization can overcome the attractive force in the ionic potential. In this case, the Cs$^+$ ion is liberated from the droplet and detected. The second possibility is that the kinetic energy of the produced ion is not sufficient for the ion to escape from the He nanodroplet. In this case, the ion falls back into the droplet and is not detected. 
Its kinetic energy is dissipated as a result of its interaction with the droplet during the solvation process. The critical value of the time delay between the excitation and ionization steps, which separates the desorption from the fall back and solvation of Cs$^+$, is usually termed the fall-back time.

In the following, experimental results on the dynamics of Cs atoms attached to He nanodroplets and photoexcited to the three molecular states depicted in Figure \ref{fig:fallback schematic+levels scheme}b will be presented. Two different detection techniques, velocity map imaging (VMI) and ion-time-of-flight (ion-TOF) mass spectrometry, were used. The ion-TOF measurements have allowed us to assign the masses of the detected ions. Electron- and ion-VMI detection have allowed us to extract the relative state populations and the kinetic energies of the detected ions. Furthermore, VMI detection provides information about possible anisotropies in the ion emission.
\section{Experimental Section}
\label{sec:1}
\begin{figure}
\includegraphics[width=\columnwidth]{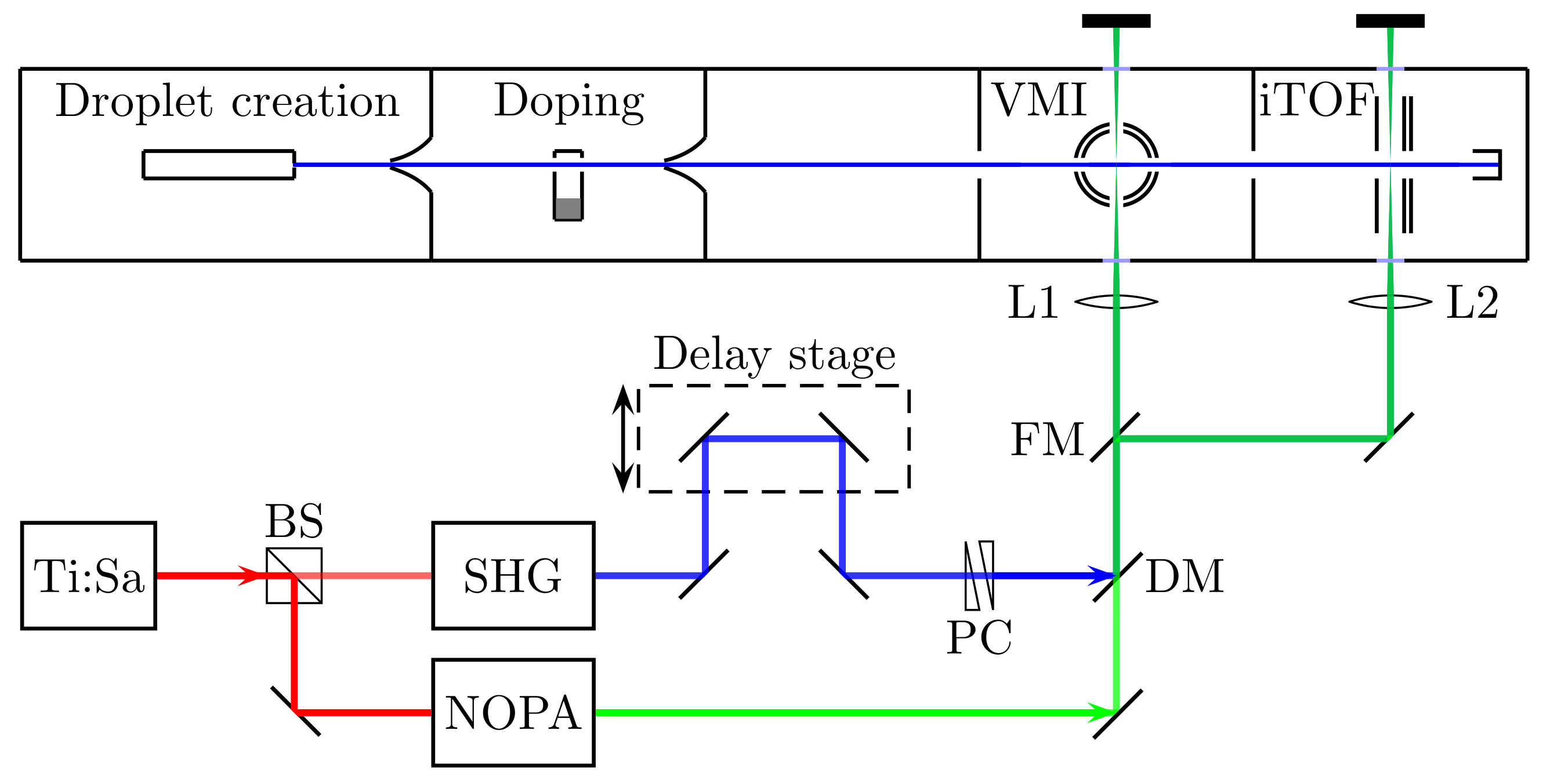}
\caption{Schematic drawing of the experimental setup. Abbreviations: Ti:Sa = titanium:sapphire laser system, BS = nonpolarizing beamsplitter, SHG = $\beta$-barium borate crystal for second harmonic generation, NOPA = nonlinear optical parametric amplifier, PC = prism compressor, DM = dichroic mirror, FM = flip mirror, L1 and L2 = lenses with a focal length of 75\,cm each, VMI = velocity map imaging spectrometer, iTOF = ion time-of-flight mass spectrometer.}
\label{fig:setup}       
\end{figure}

Figure \ref{fig:setup} shows a schematic drawing of the setup used for the experiments presented here. In brief, a beam of He nanodroplets with a mean size ranging from 1000 to 10000 atoms per droplet is generated by a continuous supersonic expansion of He through a nozzle with a 5\,$\mu$m opening diameter. The central part of the beam passes through a 0.4\,mm diameter skimmer placed 13\,mm downstream from the nozzle, which reduces the transverse velocity spread of the beam and allows for a downstream propagation of the supersonic beam under high-vacuum conditions. Thereafter, the beam enters a vacuum chamber which contains a radiatively-heated pick-up cell held at 353\,K. These temperature conditions ensure that single Cs atom doping of the He nanodroplets is predominant. The Cs-doped He nanodroplets then intersect with the pump and probe laser pulses in vacuum chambers which contain a VMI detector and an ion-TOF mass spectrometer, respectively. The VMI detector used here, which is oriented in a direction perpendicular with respect to the He nanodroplet beam, is described in detail in ref.\,\cite{Fechner.2012}. To record the velocity-map images, a fast CMOS camera (acA1920-155um - Basler ace) and a centroiding computer algorithm are used. The electron- and ion-velocity-map distributions are reconstructed by using the maximum entropy velocity map reconstruction method\,\cite{Dick.2014}. The ion-TOF spectrometer is mounted in-line with the He nanodroplet beam. It includes two high-voltage extraction electrodes, a field-free drift region and a Daly-type detector\,\cite{Daly.1960}, which is composed of a Faraday cup, a scintillator plate, an optical bandpass filter to block stray laser light and a photomultiplier tube (PMT). The ion-TOF signals are recorded using a fast digitizer card (Acqiris DP105/U1067A-001).

The femtosecond excitation (pump) pulses at center wavenumbers $E_{\mathrm{pump}}/(hc)$ between 11120 and 11990 cm$^{-1}$ (where $h$ is Planck's constant and $c$ is the speed of light; see Figure \ref{fig:fallback schematic+levels scheme})
% wavelengths of $\lambda_1$ = \SIlist[list-units=single,list-final-separator = {, }, list-pair-separator= {, }] {834;844;851;859;888;899}{\nano\metre}
are produced by feeding a part of the output of a regenerative Titanium:Sapphire (Ti:Sa) amplifier ($E/(hc)$ = 12500 cm$^{-1}$) into a nonlinear optical parametric amplifier (NOPA). For the ionization (probe) step, we produce frequency-doubled femtosecond laser pulses at $E_{\mathrm{probe}}/(hc) = 25000$ cm$^{-1}$ by sending a portion of the laser light of the regenerative amplifier through a $\beta$-barium borate (BBO) crystal. The 5\,kHz repetition rate of the regenerative amplifier assures a sufficiently large period between subsequent pulse pairs to prevent multiple photoexcitations of the alkali atoms. The time delay between the excitation pulse and the ionization pulse is scanned using a mechanical delay stage before the pulses are recombined collinearly with a dichroic mirror. The duration of the laser pulses is in between $100 \leq t_\mathrm{p} \leq 200$\,fs depending on the used center wavenumber. The pulses are focused into the interaction region by using a lens (75\,cm focal length) resulting in intensities of $\approx\, 10^{11}$\,W/cm$^{2}$ for the pump pulses and $\approx\, 10^{13}$\,W/cm$^{2}$ for the probe pulses. The beam paths FM-L1 and FM-L2 shown in Figure \ref{fig:setup} are used for the VMI and the iTOF measurements, respectively. Signal contributions produced by the pump or the probe pulses only are subtracted from the pump-probe-correlated signal. The polarizations of the pump and probe laser beams are set perpendicular to the VMI spectrometer axis. 
\section{Results and Discussion}
\begin{figure*}
\includegraphics[width=1.6\columnwidth]{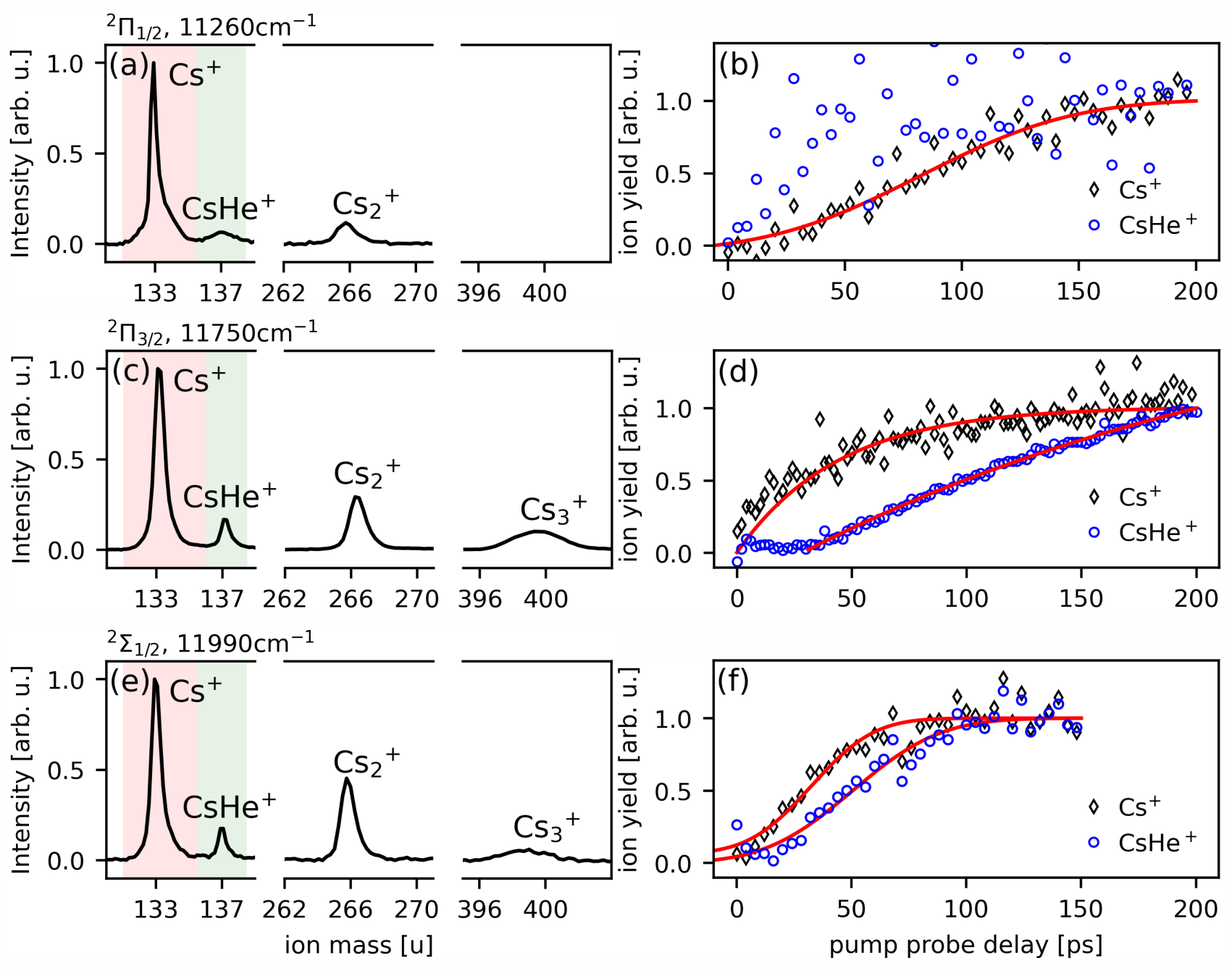}
\caption{Results obtained using laser excitation of the 6p $^{2}\Pi _{1/2}$, 6p $^{2}\Pi _{3/2}$ and $^{2}\Sigma _{1/2}$ states at the wavenumbers $E_{\mathrm{pump}}/(hc)$ given in the figure, followed by laser ionization at $E_{\mathrm{probe}}/(hc) = 25000$ cm$^{-1}$ and subsequent ion-TOF detection.
(a, c, e) Exemplary ion-TOF mass spectra obtained at pump-probe delays of 200\,ps, 200\,ps and 150\,ps, respectively. (b, d, f) Integrated Cs$^+$ and CsHe$^{+}$ yields obtained by integration of the ion-TOF traces in the intervals marked by the red and green shadings in (a, c, e) and subsequent smoothing as a function of pump-probe delay, respectively. The experimental data obtained from one measurement are shown as black and blue markers, respectively. The solid red lines are obtained from fits to the experimental data (see the main text). The ion yield data and fit curves are normalized to the values of the respective fit curves at the largest pump-probe delay.}

\label{fig:allionTOFplots} 
\end{figure*}
As can be seen from Figures \ref{fig:allionTOFplots}a,c,e, most ion-TOF spectra contain contributions from Cs$^+$, CsHe$^+$, Cs$_2^+$ and Cs$_3^+$. For Cs$_2^+$ (Cs$_3^+$) resulting from the desorption of Cs$_2$ (Cs$_3$) and subsequent photo-ionization of the molecule, we did (did not) observe delay-dependent dynamics at all excitation wavenumbers studied here. Because there is currently no theory work available which would allow us to describe the desorption dynamics of Cs$_x$ (where $x = 2, 3, ...$), only the results obtained for the desorption of photoexcited Cs atoms and CsHe exciplexes are discussed below.
For all ion-VMI measurements presented in this work, no pronounced anisotropies have been found.

We have also conducted experiments at different doping conditions by varying the temperature of the pick-up cell in the range of 328--348 K. This also leads to a shrinking of droplet sizes as a result of evaporative cooling which arises from the energy deposition upon pick-up. At the different doping conditions studied here, we did not observe a change in the desorption dynamics. This behavior can be attributed to the local character of the interaction of the Cs atom with the He nanodroplet. This finding is in agreement with previous experimental and theory work on other alkali-doped He nanodroplets, such as Na-He$_{N}$\,\cite{Hernando.2012}.
In the following, the subscripts ``V'' and ``T'' are used to distinguish in between results obtained by using VMI and ion-TOF detection, respectively. The superscript ``e'' (no superscript) is used to label results obtained by using electron-VMI (ion-VMI and ion-TOF) detection. The subscript ``c'' denotes calculated results from theory work. Because two different functional forms are used to fit the experimental results, the subscript ``R'' is used to label the results obtained from a growth function (see below).
An overview of all the experimental results obtained in this work and the corresponding theory values is given in Tab.\,\ref{tab:1}.

\subsection{Desorption and Solvation of Cs Atoms Excited to the 6p $^2$\textPi$_{1/2}$ State}
\label{section:pi one half}

\begin{figure*}
\includegraphics[width=1.6\columnwidth]{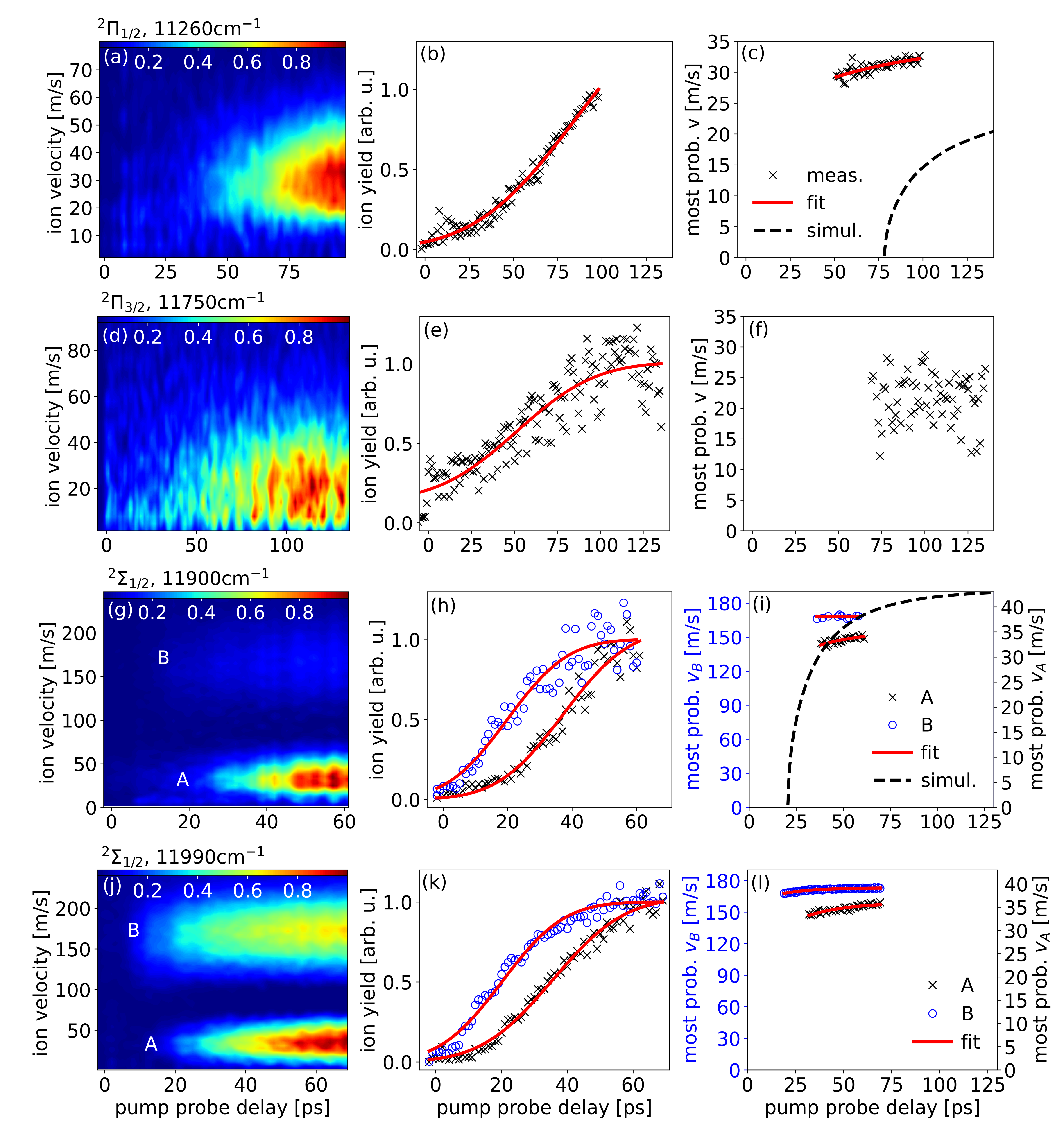}
\caption{(a, d, g, j) Normalized ion velocity distributions, (b, e, h, k) normalized ion yields, and (c, f, i, l) most probable ion velocities as a function of pump-probe delay obtained by using ion-VMI detection. Each row  corresponds to a selected wavenumber $E_{\mathrm{pump}}/(hc)$ (given in the figure). The wavenumber for the probe pulse was $E_{\mathrm{probe}}/(hc) = 25000\,cm^{-1}$. The experimental data are shown as filled contour lines and as black and blue markers. The solid red lines are the results of fits to the experimental data (see the main text). In (g--l), two features A and B are observed which exhibit different delay-dependent characteristics. The fit curve shown in (i) is for feature A only. The data used for the dashed black lines in (c) and (i) are extracted from ref.\,\cite{Coppens.2019}. The intensity scales in (a, d, g, j) range from 0 to 1. The ion yield data and fit curves are normalized to the values of the respective fit curves at the largest pump-probe delay.}
\label{fig:allionVMIplots} 
\end{figure*}

Figures \ref{fig:allionVMIplots}a--c show the ion yields and ion velocities as a function of pump-probe delay which were obtained by resonant excitation of the 6p $^{2}\Pi _{1/2}$ state at $E_{\mathrm{pump}}/(hc) = 11260$ cm$^{-1}$ and subsequent ionization at $E_{\mathrm{probe}}/(hc) = 25000$ cm$^{-1}$ using VMI detection. In all of the ion-VMI measurements presented in this article, the arrival times of the ions were not gated due to the low overall ion signal intensity. From the results of the ion-TOF measurements discussed in more detail below, we can infer that for excitation to the 6p $^2$\textPi$_{1/2}$ state the ion signal intensity in the VMI traces is mainly from Cs$^+$. Therefore, the time dependence shown in Figures \ref{fig:allionVMIplots}a--c can be ascribed mainly to Cs$^+$. Only a very weak signal intensity arising from CsHe exciplexes is recorded in the ion-TOF measurements, which does not allow for a measurement of its dynamics. This observation is in agreement with the dynamics expected from the calculated diatomic potential energy curve for CsHe exciplexes in the 6p $^2\Pi_{1/2}$ state, which has a barrier that leads to a strong suppression of exciplex formation\,\cite{Coppens.2019}. Exciplex formation was also not reported for Cs-doped He nanodroplets after excitation to the 6p $^2\Pi_{1/2}$ state\,\cite{Theisen.2011} and for Rb-doped He nanodroplets after excitation to the 5p $^2\Pi_{1/2}$ state\,\cite{Theisen.2011b, Aubock.2008}. 

The time-dependent ion yield shown in Figure \ref{fig:allionVMIplots}b was fitted to an error function of the form
\begin{equation}
\emph{I}(\emph{t}) = A_0 \{ \, \mathrm{erf}[\left(t-\tau_{\mathrm{V}}\right) /\sigma_{\mathrm{V}}]+1\},
\label{erfun}
\end{equation}
with variable amplitude $A_0$, fall-back time constant $\tau_{\mathrm{V}}$ and width $\sigma_{\mathrm{V}}$. This functional form, which has also previously been used to describe the desorption dynamics of alkali atoms from He nanodroplets \cite{Vangerow.2015}, contains an exponential term which fits to the characteristics of a desorption process. The fit yields a fall-back time of $\tau_{\mathrm{V}}$ = 72 ps and a width of $\sigma_{\mathrm{V}}$ = 44\,ps. In Figure\,\ref{fig:allionVMIplots}\,c, the most probable ion velocity is plotted as a function of pump-probe delay. This velocity is determined by fitting the velocity distribution at each delay step (shown in Figure \ref{fig:allionVMIplots}a) with a Gaussian function. The most probable ion velocity at each pump-probe delay is then given by the peak value of the Gaussian fit to the ion velocity distribution. At pump-probe delays, where the ion signal intensity is very low, and the corresponding fits yield very noisy or unphysical values for the extracted velocities, the results are not used for the further analysis, and they are thus not shown in Figure \ref{fig:allionVMIplots}. The asymptotic ion velocity, $v_{\mathrm{V}}$, which is the maximum ion velocity reached at an infinitely long pump-probe delay, is obtained by fitting the delay-dependent most probable ion velocities in Figure \ref{fig:allionVMIplots}c to a function of the form
\begin{equation}
\gamma(t) = v_{\mathrm{V}}\{ 1-\exp[{-(t-t_0)/s_{\mathrm{V}}}]\},
\label{v_asymp}
\end{equation}
with the asymptotic ion velocity $v_{\mathrm{V}}$, the time offset $t_0$, and width $s_{\mathrm{V}}$. At this excitation wavenumber, we obtain an asymptotic ion velocity $v_{\mathrm{V}}$ = 35\,m/s and a width $s_{\mathrm{V}}$ $<$ 1\,m/s.

The obtained fall-back times and asymptotic ion velocities are close to the theory values of $\tau_{\mathrm{c}}$ =77.8\,ps and $v_{\mathrm{c}}$ = 23.8\,m/s given by Coppens et al.\,\cite{Coppens.2019} which were derived by using time-dependent $^4$He-DFT simulations and attributed to to impulsive dissociation, which is expected from the corresponding repulsive He-Cs(6p) interaction potential.
The calculated time-dependent velocities, extracted from ref.\cite{Coppens.2019}\,, are included as dashed lines in Figures \ref{fig:allionVMIplots}c,i. In comparison, the most probable velocities obtained from the measurements miss the steep increase starting at the fall-back time which is apparent from the theory curves, and they display high velocity values close to or even below the nominal fall-back times. We attribute this behavior to the broad distribution of excitation energies in combination with the steep repulsive potential. The calculations do not include the distribution of excitation energies. In the experiment, around the fall-back time the average ion intensity is low; on the other hand, the blue tail provides already contributions with high ion velocity.

The  measurements were repeated at a slightly red-detuned excitation wavenumber of $E_{\mathrm{pump}}/(hc) = 11120$ cm$^{-1}$. Even though the recorded number of photoelectrons is comparable to the number of photoelectrons obtained for the resonant excitation of this state at $E_{\mathrm{pump}}/(hc) = 11260$ cm$^{-1}$, we were unable to detect ions under these excitation conditions in the VMI measurements. 
These observations are consistent with previous results by Theisen et al.\,\cite{Theisen.2011} who explained these findings by the nondesorptive nature of this state when excitation occurs at the lower energy tail of its dipole absorption spectrum. 
The nondesorptive character is due to the less strong repulsion of the excited Cs atom from the droplet at this excitation wavenumber, which does not allow the excited Cs atom to gain enough kinetic energy to leave the Cs-He$_N$ potential before the interaction is switched to attraction in the ionization step.

However, we were able to detect a small amount of Cs$^+$ in the corresponding ion-TOF measurements at an excitation wavenumber of $E_{\mathrm{pump}}/(hc) = 11120$ cm$^{-1}$. This can be explained by some overlap of the broad laser excitation spectrum (full width at half-maximum  $\mathrm{(FWHM)}\approx149\,\mathrm{cm}^{-1}$) with the high-energy tail of the absorption profile for this state. Because of the different sensitivities of the ion-VMI and ion-TOF detection schemes, weak ion signal contributions may not show up in the VMI measurements.

\begin{table*}
\footnotesize
\begin{center}
\caption[]{Summary of fall-back times $\tau_{\mathrm{V}}$ and $\tau_{\mathrm{T}}$ inferred from the ion-VMI and from the ion-TOF measurements, respectively, at different central excitation wavenumbers $E_{\mathrm{pump}}/(hc)$. The full width at half-maximum (FWHM) bandwidth of each laser excitation spectrum is provided in brackets behind the respective excitation wavenumber. The widths $\sigma_{\mathrm{V}}$ and $\sigma_{\mathrm{T}}$ obtained from the corresponding fits of the ion yields to error functions are given in parantheses. Likewise, the obtained asymptotic ion velocities $v_{\mathrm{V}}$ as well as the 50\,\% rise times $\tau_{\mathrm{T,R}}$ and $\tau^\mathrm{e}_{\mathrm{V,R}}$ from the ion-TOF and from the electron-VMI measurements are given, respectively. The uncertainty of each asymptotic ion velocity $s_{\mathrm{V}}$ (given in parantheses) is quoted either as the error of the corresponding fit to Eq.\,\ref{v_asymp} or as the standard deviation of the most probable ion velocities (see the main text).\footnote{In several VMI traces, two features (labeled as ``A'' ,``B'' for ions and ``I'', ``II'' for electrons, respectively) with different time constants were found to be indicative for desorption processes. These time constants are thus given separately. In the ion-TOF measurements, the Cs$^+$ and CsHe$^+$ ion masses were found to exhibit different delay dependencies. Therefore, $\tau_{\mathrm{T}}$ and $\tau_{\mathrm{T,R}}$ are given for both ion masses separately. 
The calculated fall-back times $\tau_{\mathrm{c}}$ and the asymptotic ion velocities $v_{\mathrm{c}}$ for Cs$^+$ obtained by Coppens et al.\,\cite{Coppens.2019} are shown for comparison. 
For the measurement at $E_{\mathrm{pump}}/(hc) = 11120$\,cm$^{-1}$ ($E_{\mathrm{pump}}/(hc) = 11640$\,cm$^{-1}$), no fall-back times and asymptotic ion velocities are given, since desorption dynamics was not observed (the signal intensity was not sufficient).}}
\label{tab:1}
\begin{tabular*}{\textwidth}{c@{\extracolsep{\fill}}ccccccccc}
\hline
\noalign{\smallskip}
state & 
$E_{\mathrm{pump}}/(hc)$ (FWHM)\,[cm$^{-1}$] 
& 
label & 
$\tau_{\mathrm{V}}$($\sigma_{\mathrm{V}}$)\,[ps] &
$\tau_{\mathrm{T}}$($\sigma_{\mathrm{T}}$)\,[ps] &
$\tau_{\mathrm{T,R}}$\,[ps] &
$\tau^\mathrm{e}_{\mathrm{V,R}}$\,[ps] &
$\tau_{\mathrm{c}}$\,[ps] &
$v_{\mathrm{V}}$($s_{\mathrm{V}}$)\,[m/s] &
$v_{\mathrm{c}}$\,[m/s] \\

\noalign{\smallskip}\hline\noalign{\smallskip}
6p $^2$\textPi$_{1/2}$ & 11260 (152) & & 72(44) & Cs$^{+}$:\,82(73) & & & 77.8 & 35($<$1) & 23.8\\
6p $^2$\textPi$_{1/2}$ & 11120 (149) & & & & & & & & \\
6p $^2$\textPi$_{3/2}$ & 11750 (166) & & 55(51) & & Cs$^+$:\,29 & & & 22(4) & \\
& & & & & CsHe$^{+}$:\,152 & & & & \\
6p $^2$\textPi$_{3/2}$ & 11640 (163) & & & & & & & & \\
6p $^2$\textSigma$_{1/2}$ & 11900 (170) & A, I & 36(19) & & & 21 & 20.2 & 36($<$1) & 43.6\\
 & & B, II & 16(26) & & & & & 169(1) & \\
6p $^2$\textSigma$_{1/2}$ & 11990 (173) & A, I & 35(26) & Cs$^{+}$:\,30(27) & & 23 & & 36(3) & \\
 & & & & CsHe$^+$:\,49(41) & & & & & \\
 & & B, II & 20(17) & & & & & 173(3) & \\
\noalign{\smallskip}
\hline
\end{tabular*}
\end{center}
\end{table*}

The delay-dependent Cs$^+$ ion yields from complementary ion-TOF measurements, following excitation to the 6p $^2\Pi_{1/2}$ state, were obtained by using the same experimental parameters as in the VMI measurements. We only observe electrons and no bare Cs$^+$ ions and assign the missing positive charge to He droplet attached Cs$^+$ ions, which we cannot detect because of the very large mass. In this way the results confirm the solvation of Cs$^+$ inside the He droplet upon excitation at $E_{\mathrm{pump}}/(hc) < 11120$\,cm$^{-1}$.

At an excitation wavenumber of $E_{\mathrm{pump}}/(hc) = 11260$\,cm$^{-1}$, the Cs$^+$ ions were found to desorb with a fall-back time of $\tau_{\mathrm{T}}=$ 82\,ps. 
This time constant was obtained from a fit of the experimental Cs$^+$ ion data (see Figure \ref{fig:allionTOFplots}b) to the error function given by eq.\,\ref{erfun}. The corresponding width of the error function was found to be $\sigma_{\mathrm{T}} =$ 73\,ps.
These values are fairly consistent with the results obtained from the ion-VMI measurements.

\subsection{Dynamics of Cs Atoms Excited to the 6p $^2$\textPi$_{3/2}$ State}
\begin{figure}
\includegraphics[width=0.9\columnwidth]{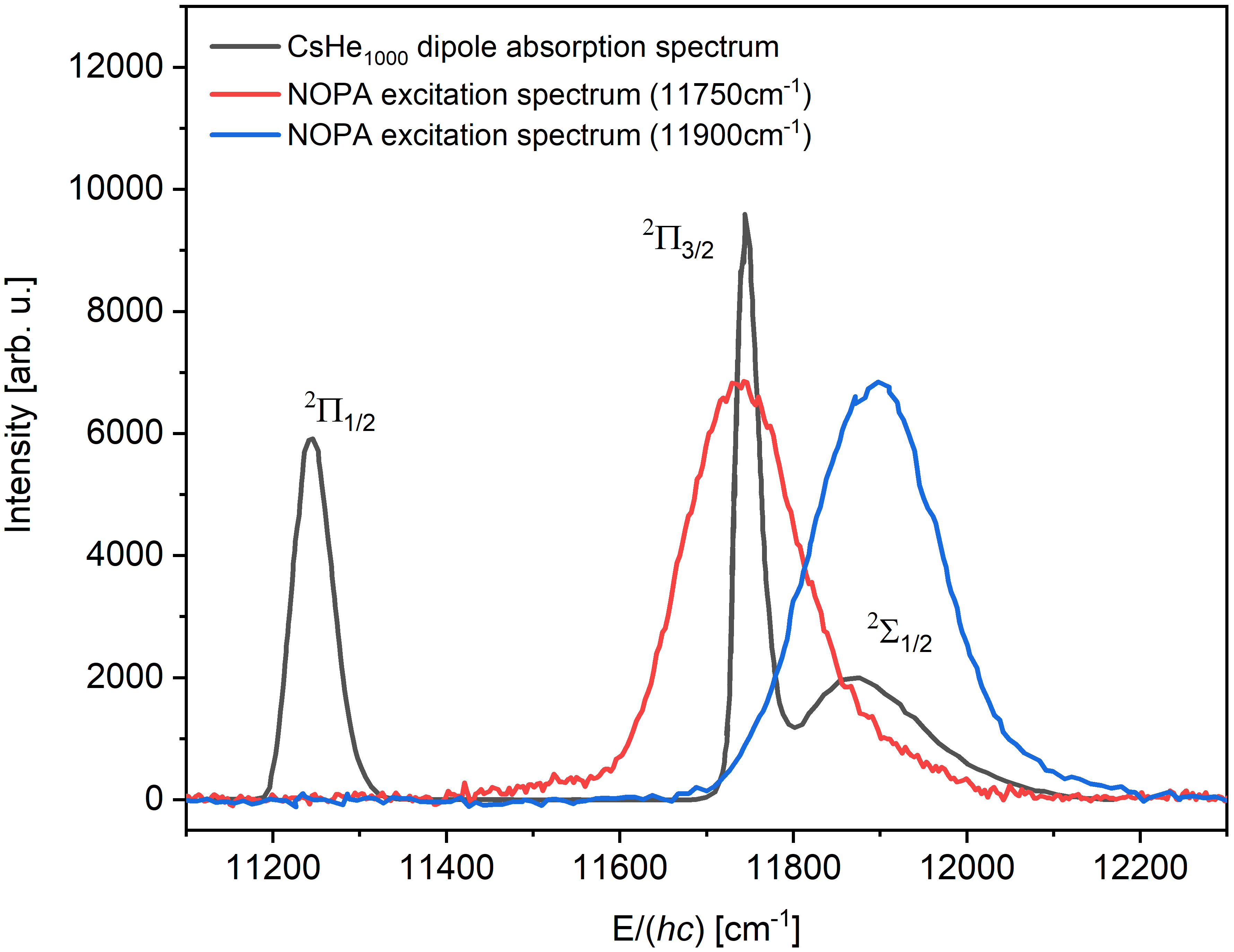}
\caption{(black, solid): Calculated dipole absorption spectrum of a Cs-He$_N$ complex containing $N = 1000$ $^4$He atoms (black curve) and two measured NOPA excitation spectra at different center wavenumbers (blue and red curves, see legend). The NOPA spectra were normalized to their corresponding maxima. The data for the dipole absorption spectrum were obtained from ref.\,\cite{Coppens.2019}}.
\label{fig:dipole absorption} 
\end{figure}
The results of the time-dependent $^4$He-DFT simulations for the excitation of the 6p $^2\Pi_{3/2}$ state predict that the excited Cs atom quickly forms a linear exciplex molecule which does not desorb from the He nanodroplet but stays at the droplet surface\,\cite{Coppens.2019}. % OLD: predict that the excited Cs atoms do not desorb from the He nanodroplet, but stay at the droplet surface\,\cite{Coppens.2019}.
Figure \ref{fig:allionVMIplots}d--f show the results of the ion-VMI measurements obtained upon resonant excitation of this state at $E_{\mathrm{pump}}/(hc) = 11750$ cm$^{-1}$. In Figure \ref{fig:dipole absorption}, the measured spectra of the broad-band NOPA laser are plotted together with the dipole absorption spectrum of a CsHe$_N$ complex.
Because of the large width of the laser excitation spectrum ($\mathrm{FWHM}\approx166\,$cm$^{-1}$), it is expected that the 6p $^2\Sigma_{1/2}$ state is also populated during the excitation step. However, the results from the ion-VMI measurements imply that the excited Cs atoms \textit{do} desorb from the He nanodroplet. The fall-back time of the ions and the corresponding width (obtained by fitting the experimental ion traces to eq.\,\ref{erfun}) are $\tau_{\mathrm{V}}=$ 55\,ps and $\sigma_{\mathrm{V}}=$ 51\,ps, respectively. The asymptotic ion velocity, which was determined to be the mean value of the most probable ion velocities shown in Figure \ref{fig:allionVMIplots}d, is $v_{\mathrm{V}}$ = 22\,m/s (standard deviation of $s_{\mathrm{V}}$ = 4\,m/s).

From the delay-dependent ion-TOF yields in Figure \ref{fig:allionTOFplots}d, it can be inferred that both excited-state Cs atoms and CsHe exciplexes appear to desorb from the He nanodroplets. At long pump-probe delays, the detected number of Cs$^+$ ions seems to have reached a constant value which suggests that all excited Cs atoms have desorbed. In contrast to that, the CsHe$^+$ yield is still increasing at the longest measured delays.

Because fits to Eq.\,\ref{erfun} did not yield satisfactory results, the Cs$^+$ and CsHe$^+$ ion yields (see Figure \ref{fig:allionTOFplots}d) are fit to the following functional form
\begin{equation}
\tilde{I}(t) = \tilde{A}_0\{ 1-\exp[{-\mathrm{\ln\left(2\right)}(t-t_0)/\tau}]\},
\label{fiftypercentfun}
\end{equation}
with variable amplitude $\tilde{A}_0$, offset $t_0$ and width $\tau$. The fits yield 50\,$\%$-rise times of $\tau_{\mathrm{T,R}} = t_0 + \tau =$ 29\,ps and $\tau_{\mathrm{T,R}} = t_0 + \tau =$ 152\,ps for Cs$^+$ and CsHe$^+$, respectively. The fit to the CsHe$^+$ yield was started at a nonzero pump-probe delay of $t_0 =$ 30\,ps to account for the delayed appearance of the CsHe$^+$ signal in the ion-TOF measurements. This value is fairly consistent with the predicted CsHe exciplex formation time of 25\,ps\,\cite{Coppens.2019}. Even though desorption is predicted not to occur for bare Cs atoms after resonant excitation of the 6p $^2\Pi_{3/2}$ state\,\cite{Coppens.2019}, we do observe a considerable amount of Cs$^+$ ions (see Figure \ref{fig:allionTOFplots}c). A small fraction of this Cs$^+$ yield is expected to be due to a parasitic population of the 6p $^2\Sigma_{1/2}$ state due to the overlap of the laser excitation spectrum with the lower energy tail of the dipole absorption peak corresponding to the 6p $^2\Sigma_{1/2} \leftarrow$ 6s $^2\Sigma_{1/2}$ transition (see Figure \ref{fig:dipole absorption}). However, the main signal contribution is expected to arise from the excitation of the narrow and much stronger 6p $^2\Pi_{3/2} \leftarrow$ 6s $^2\Sigma_{1/2}$ transition. Furthermore, a comparison of the Cs$^+$ yields from ion-TOF spectrometry in Figure \ref{fig:allionTOFplots} e (obtained by resonant excitation of the 6p $^2\Sigma_{1/2}$ state) with those shown in Figure \ref{fig:allionTOFplots}c (obtained by resonant excitation of the 6p $^2\Pi_{3/2}$ state) suggests that the involved solvation dynamics of the excited Cs atoms are different. 
In contrast to the other excitation schemes, for which error functions were used for fitting the experimental data, the delay-dependent increase of the Cs$^+$ ion yield in Figure \ref{fig:allionTOFplots}d is described best by using the growth function given by eq.\,\ref{fiftypercentfun}. In addition to that, the ion velocity distribution shown in Figure \ref{fig:allionVMIplots}d appears to be broader compared to the distributions observed for the excitation of the other states in this work. These findings indicate that the desorption could be driven by a more complex, evaporative-like process which deviates from the impulsive, dissociation-like desorption of Cs atoms in the 6p $^2\Pi_{1/2}$ and 6p $^2\Sigma_{1/2}$ states. Such evaporative-like desorption dynamics were suggested for the 5p states of Rb which were observed to exhibit much slower desorption dynamics compared to other states\,\cite{Vangerow.2017, Coppens.2018}.

As proposed by Coppens et al.\,\cite{Coppens.2019}, CsHe exciplexes can also form on the surface of the droplet after excitation of the 6p $^2\Pi_{3/2}$ state, since there is no barrier in the calculated diatomic potential of the CsHe exciplex in this case. After exciplex formation, spin-orbit relaxation to the 6p $^2\Pi_{1/2}$ state may occur which could lead to the desorption of CsHe exciplexes or of bare Cs atoms. The contribution of spin-orbit relaxation to desorption depends on how the energy, which is released by this relaxation process, is shared in between the He nanodroplet and the CsHe exciplex. 
Because we do not detect photoelectrons with an energy corresponding to the 6p $^2\Pi_{1/2}$ state, this relaxation process does not seem to occur. Vibrational predissociation could also lead to the desorption of CsHe exciplexes, since the vibrational energy of the CsHe exciplexes is sufficient to release CsHe from the He nanodroplet. This process would be similar to the vibrational predissociation of rare gas-diatomic molecule van der Waals complexes reported in the literature\,\cite{Beswick.1977, Beswick.1981a, Rohrbacher.2000, Beswick.2012}.
% OLD: {Beswick.1980, Beswick.1981, GonzalezLezana.1996}.
The results of a theoretical study by Leino et al., which addresses the formation and possible desorption of RbHe exciplexes in the 5p $^2\Pi_{3/2}$ state, suggest that the internal binding energy of the RbHe exciplex is approximately 10 times higher than the binding energy of the RbHe exciplex to the He nanodroplet\,\cite{Leino.2011}. 

To evaluate whether vibrational predissociation could be a possible process leading to the desorption of CsHe exciplexes after excitation of the 6p $^2\Pi_{3/2}$ state, we have fitted the pseudo-diatomic potential energy curve of the 6p $^2\Pi_{3/2}$ state, given in ref.\,\cite{Coppens.2019}, to a Morse potential function, from which the vibrational energy eigenvalues of the Cs(6p)-He$_N$ complex can be calculated analytically. Using this procedure, we obtain four bound vibrational levels with principal quantum numbers $n=0,1,2$ and $3$ and the corresponding binding energies $E_n$ = $-88.0$\,cm$^{-1}\cdot hc$, $-46.6$\,cm$^{-1}\cdot hc$, $-18.2$\,cm$^{-1}\cdot hc$, $-2.9$\,cm$^{-1}\cdot hc$, respectively.
It is likely that the binding energy of CsHe exciplexes to the He nanodroplet is close to the estimated binding energy of $\approx$ 10\,cm$^{-1}\cdot hc$ for RbHe exciplexes\,\cite{Leino.2011}.
Vibrational relaxation can therefore easily provide enough energy to break the bond between the CsHe exciplex and the He nanodroplet which supports the above-mentioned explanation for the desorption of CsHe exciplexes. Furthermore, the delay-dependent ion-TOF yield for CsHe$^+$ obtained from this measurement follows a different time dependence as compared to the other CsHe$^+$ ion-yield curves shown in this work. This points to a desorption process which is different from impulsive desorption, for which the delay-dependent ion yield usually follows an error function. Even though our results provide strong evidence for CsHe exciplex desorption as a result of vibrational predissociation, more experimental and theoretical efforts are required to allow for a confident conclusion on the process that causes the desorption of CsHe exciplexes after resonant excitation to the 6p $^2\Pi_{3/2}$ state.

Because VMI detection could not be used for mass-selected ion spectrometry, the velocity at which the CsHe exciplexes desorb from the droplet is not known. Judging from the relative ion yields obtained from the ion-TOF measurements, it is likely that the delay-dependent dynamics observed in the ion-VMI data arise from both excited Cs atoms and CsHe exciplexes. This could also explain the different fall-back times obtained using ion-VMI and ion-TOF detection (see Tab.\ \ref{tab:1}). Since the ion-TOF data were recorded in a mass-selected manner, the fall-back times for Cs$^+$ obtained from the ion-TOF measurements are more reliable than the fall-back times obtained from the ion-VMI measurements.
\subsection{Dynamics of Cs Atoms Excited to the 6p $^2$\textSigma$_{1/2}$ State}
Because of the broad excitation spectrum of the NOPA (width of $\mathrm{FWHM}\approx166$\,cm$^{-1}$) and due to the closely-spaced absorption bands for excitation to the 6p $^{2}\Sigma _{1/2}$ and 6p $^{2}\Pi _{3/2}$ states (see Figure \ref{fig:dipole absorption}), VMI measurements were done at two different laser excitation wavenumbers to determine the dynamics of the 6p $^{2}\Sigma _{1/2}$ state. In the first measurement, the wavenumber of the excitation laser is centered near the maximum absorption cross section for the 6p $^{2}\Sigma _{1/2}$ state ($E_{\mathrm{pump}}/(hc) = 11900$\,cm$^{-1}$). In the second measurement, the wavenumber of the excitation laser is blue-detuned from the resonance ($E_{\mathrm{pump}}/(hc) = 11990$\,cm$^{-1}$) to ensure that the contribution of the 6p $^{2}\Pi _{3/2}$ state to the signal intensity is negligible.

\begin{figure*}
\includegraphics[width=1.6\columnwidth]{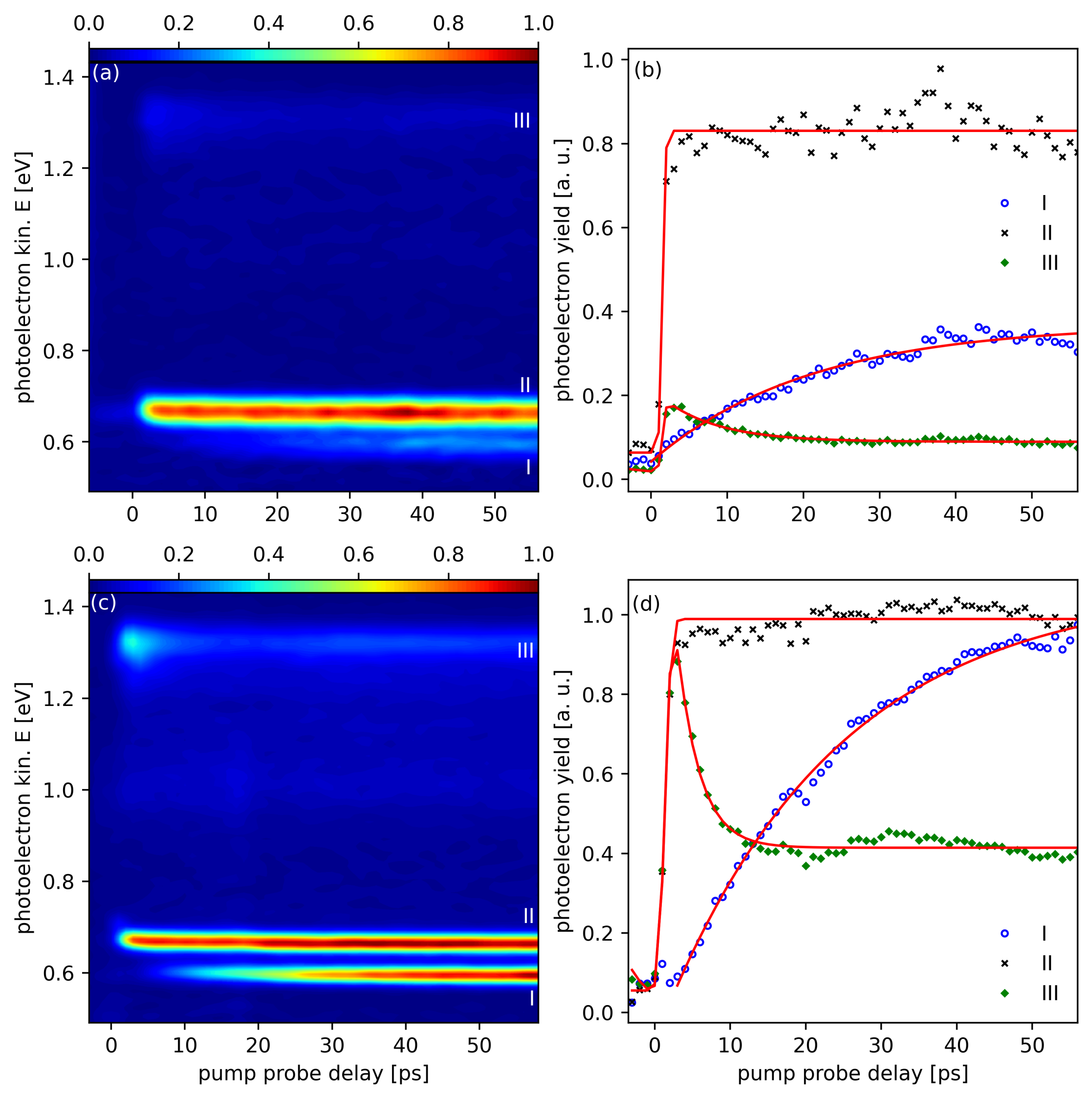}
\caption{Photoelectron kinetic energy distributions resulting from electron-VMI detection following laser excitation of the 6p $^{2}\Sigma _{1/2}$ state at (a) $E_{\mathrm{pump}}/(hc) = 11900$\,cm$^{-1}$ and (c) $E_{\mathrm{pump}}/(hc) = 11990$\,cm$^{-1}$ and subsequent laser ionization at $E_{\mathrm{probe}}/(hc) = 25000$\,cm$^{-1}$ at different pump-probe delays. (b), (d) Delay-dependent photoelectron yields for features I (blue markers), II (black markers) and III (green markers) obtained by integrating the photoelectron kinetic energy distributions shown in (a) and (b) in the corresponding feature ranges, respectively. The red curves were obtained from fits to the delay-dependent photoelectron yields (see the main text).}
\label{fig:allelectronVMIplots} 
\end{figure*}
Figure \ref{fig:allelectronVMIplots}a shows the delay-dependent photoelectron kinetic energy distributions which were obtained by using electron-VMI detection at an excitation wavenumber of $E_{\mathrm{pump}}/(hc) = 11900$\,cm$^{-1}$. As can be seen from the figure, three different features are detected which are labeled as I, II and III. 
The photoelectrons for feature I have an energy of 0.60\,eV, which can be assigned to ionization out of the 6p $^{2}\Pi _{1/2}$ state. The photoelectrons obtained for feature II have a mean kinetic energy of 0.66\,eV which corresponds to the electron energy resulting from the ionization out of the 6p $^{2}\Sigma _{1/2}$ and 6p $^{2}\Pi _{3/2}$ states at $E_{\mathrm{probe}}/(hc) = 25000$\,cm$^{-1}$. 
At this excitation wavenumber, the number of detected electrons for feature I is lower than for feature II and the electrons for feature I appear to occur at a time delay compared to those for feature II. The $50\,\%$ rise time for feature I, which is obtained by fitting the corresponding photoelectron yield to Eq.\,\ref{fiftypercentfun}, was determined to be $\tau^\mathrm{e}_{\mathrm{V,R}} =$ 21\,ps.

The corresponding laser excitation spectrum is well separated from the absorption band corresponding to the excitation of the 6p $^{2}\Pi _{1/2} \leftarrow$ 6s $^{2}\Sigma _{1/2}$ transition. In addition to that, the natural lifetime of the 6p state of the free Cs atom (30.41\,ns\,\cite{Young.1994}) is much longer than the maximum pump-probe delay. This suggests that the delayed population of the 6p $^{2}\Pi _{1/2}$ state arises from the perturbation of the 6p $^{2}\Sigma _{1/2}$ state by the He droplet. This perturbation leads to spin-orbit relaxation. Spin-orbit relaxation has been observed previously for other dopants attached to He droplets\,\cite{Loginov.2007, Loginov.2011, Loginov.2015, Loginov.2014, Bruhl.2001, Koch.2014, Lindebner.2014, Kautsch.2013, Vangerow.2014}.
For Cs-doped He nanodroplets, spin-orbit relaxation was proposed to occur upon excitation to the 6p $^{2}\Pi _{3/2}$ state. At this stage, it is not clear to us why spin-orbit relaxation occurs upon excitation of the 6p $^{2}\Sigma _{1/2}$ state, but not for excitation of the 6p $^{2}\Pi _{3/2}$ state. Because the atomic states are identical, spin-orbit relaxation would be expected.
However, it should be less likely for the 6p $^{2}\Sigma _{1/2}$ state than for the 6p $^{2}\Pi _{3/2}$ state considering the stronger droplet repulsion in the 6p $^{2}\Sigma _{1/2}$ state which is evident from the pseudo-diatomic potentials in the article by Coppens et al.\,\cite{Coppens.2019}.

The results of the ion-VMI measurement shown in Figures \ref{fig:allionVMIplots}g--i were obtained at the same excitation wavenumber of $E_{\mathrm{pump}}/(hc) = 11900$\,cm$^{-1}$ as the data shown in Figures \ref{fig:allelectronVMIplots}a,b. 
Two ion features A and B are observed in Figure\ref{fig:allionVMIplots}g which appear at different pump-probe delays. The ion velocity distributions are broad, but they are clearly separated from each other. The ion yield for feature B is $\approx 30\%$ of the ion yield for feature A. The corresponding pump-probe-delay dependency of the ion yield is shown in Figure \ref{fig:allionVMIplots}h.
The fall-back times $\tau_{\mathrm{V}}=$ 36\,ps and $\tau_{\mathrm{V}}=$ 16\,ps for features A and B, respectively, are obtained from fits to the error function given by eq.\,\ref{erfun}. The kinetic energies of the two ion features are significantly different. The most probable ion velocity for feature A increases from 32\,m/s at a pump-probe delay of 40\,ps to 35\,m/s at a delay of 62\,ps (Figure \ref{fig:allionVMIplots}i). Because of the low signal intensity for feature B, the most probable ion velocities (inferred from the fits to the ion velocity distribution in Figure \ref{fig:allionVMIplots}g in between pump-probe delays of 38\,ps and 60\,ps) vary in between 166\,m/s and 171\,m/s, respectively. A delay dependency could not be inferred from the most probable ion velocities for feature B in Figure \ref{fig:allionVMIplots}i. Asymptotic ion velocities for features A and B of $v_{\mathrm{V}}$ = 36\,m/s and $v_{\mathrm{V}}$ = 169\,m/s, respectively, were obtained from fits of the delay-dependent most probable ion velocities for features A and B in Figure \ref{fig:allionVMIplots}i with a function of the same form as given by Eq.\,\ref{v_asymp}. Here, the uncertainties of the most probable ion velocities for features A and B, $s_{\mathrm{V}} <$ 1\,m/s and $s_{\mathrm{V}}$ = 1\,m/s, respectively, are given as the standard deviation of the most probable ion velocity over the ten largest pump-probe delays.

The results obtained by using electron- and ion-VMI detection at an excitation wavenumber of $E_{\mathrm{pump}}/(hc) = 11990$\,cm$^{-1}$ are illustrated in Figures \ref{fig:allionVMIplots}j--l and \ref{fig:allelectronVMIplots}c,d, respectively. Owing to the little overlap of the laser excitation spectrum with the absorption band of the close-lying 6p $^{2}\Pi _{3/2}\leftarrow$ 6s $^{2}\Sigma _{1/2}$ transition, the signal contributions arising from excitation of the 6p $^{2}\Sigma _{1/2}$ state are expected to be dominant. 

At $E_{\mathrm{pump}}/(hc) = 11990$\,cm$^{-1}$, we observe that the relative signal intensities of the ion and electron features are different compared to the results obtained at $E_{\mathrm{pump}}/(hc) = 11900$\,cm$^{-1}$. The electron signals which presumably arise from spin-orbit relaxation from the 6p $^{2}\Sigma _{1/2}$ state to the 6p $^{2}\Pi _{1/2}$ state (feature I) are increased compared to the measurement at the other excitation wavenumber. At long pump-probe delays, feature I reaches almost the same yield as feature II which in turn is due to excitation via the 6p $^{2}\Sigma _{1/2}$ state. The $50\,\%$ rise time for feature I, which is obtained by fitting the corresponding photoelectron yield to Eq.\,\ref{fiftypercentfun}, was determined to be $\tau^\mathrm{e}_{\mathrm{V,R}}=$ 23\,ps. The ion yield for feature B even exceeds the ion yield for feature A. This observation indicates a clear correlation between the ions for feature B and the electrons arising from spin-orbit relaxation.

In contrast to that, the fall-back times ($\tau_{\mathrm{V}}=$ 35\,ps and $\tau _{\mathrm{V}}=$ 20\,ps) and asymptotic ion velocities ($v_{\mathrm{V}}$ = 36\,m/s and $v_{\mathrm{V}}$ = 173\,m/s) for features A and B, respectively, obtained at $E_{\mathrm{pump}}/(hc) = 11990$\,cm$^{-1}$ are similar to the values obtained at $E_{\mathrm{pump}}/(hc) = 11900$\,cm$^{-1}$.
At $E_{\mathrm{pump}}/(hc) = 11990$ cm$^{-1}$, the most probable ion velocity for feature B increases from 168\,m/s at a delay of 20\,ps to 173\,m/s at a delay of 65\,ps. For feature A, the most probable ion velocity is also observed to increase, from 33\,m/s at a delay of 31\,ps to 36\,m/s at a delay of 65\,ps. The most probable ion velocities and asymptotic ion velocities for features A and B are slightly increased when the excitation wavenumber is changed from $E_{\mathrm{pump}}/(hc) = 11900$\,cm$^{-1}$ to $E_{\mathrm{pump}}/(hc) = 11990$\,cm$^{-1}$. The observed increase of the asymptotic ion velocity at the higher excitation wavenumber is in agreement with the results reported for the desorption of other excited-state alkali atoms, e.g.\ Li, Na and Rb\,\cite{Kautsch.2013,Loginov.2014,Vangerow.2015,Vangerow.2017}. From the simulations by Coppens et al.\,\cite{Coppens.2019}, ions with a velocity of 170\,m/s are not expected to arise from a desorption mechanism caused by a repulsive pseudo-diatomic potential. Because of the obvious correlation of the signal intensities attributed to the spin-orbit relaxation from the 6p $^{2}\Sigma _{1/2}$ state to the 6p $^{2}\Pi _{1/2}$ state and the ions for feature B, we have estimated the energy release from spin-orbit relaxation in order to check whether this process can provide enough energy to explain the observed ion velocities. The energy released by this relaxation process, which was estimated from the energy difference of the potential energy curves given in ref.\,\cite{Coppens.2019}, amounts to 550\,cm$^{-1} \cdot hc$.
The energy corresponding to the most probable ion velocity for feature B is about 200\,cm$^{-1} \cdot hc$. Following this consideration, 36\,$\%$ of the released energy would be converted into the kinetic energy of the ions, and the remaining energy would be transferred to the droplet and dissipated by the evaporation of He atoms. For RbHe exciplexes, which were found to be desorbed after the relaxation from the 5p $\Pi_{3/2}$ state to the 5p $\Pi_{1/2}$ state, it is reported that 21\,$\%$ of the energy is released as kinetic energy of the ions\,\cite{JohannesvonVangerow.Mai2017}. Comparably high kinetic ion energies are reported for Ba$^+$\,\cite{Leal.2016}.
However, the authors of this latter study do not provide a final explanation for a possible process which leads to the production of such fast ions.   

The black and blue markers in Figure \ref{fig:allionTOFplots}f show data sets for the delay dependence of the Cs$^+$ and CsHe$^+$ ion yields obtained from the ion-TOF measurements at an excitation wavenumber of $E_{\mathrm{pump}}/(hc) = 11990\,$cm$^{-1}$. The corresponding fall-back times, calculated via eq.\,\ref{erfun}, are $\tau_{\mathrm{T}}=$ 30\,ps (with a width of $\sigma_{\mathrm{T}} =$ 27\,ps) for Cs$^+$ and $\tau_{\mathrm{T}}=$ 49\,ps (with a width of $\sigma_{\mathrm{T}} =$ 41\,ps) for CsHe$^+$.
The fall-back time for Cs$^+$ is again in fair agreement with the results obtained from the ion-VMI measurements and with the simulation results by Coppens et al.\,\cite{Coppens.2019}. For the excitation to the 6p $^{2}\Sigma _{1/2}$ state, the formation of CsHe exciplexes is not expected, since the corresponding Cs-He potential energy curve for this state is purely repulsive. For this reason, Coppens et al.\,\cite{Coppens.2019} do not provide fall-back times for CsHe$^+$ ions arising from the excitation of the 6p $^{2}\Sigma _{1/2}$ state. 
We attribute the observation of CsHe exciplexes in these measurements to the overlap of the laser excitation spectrum with the absorption band corresponding to the 6p $^{2}\Pi _{3/2} \leftarrow$ 6s $^{2}\Sigma _{1/2}$ transition.
The effective overlap area between the absorption band and the laser excitation spectrum provides an estimate of the expected parasitic population of the 6p $^{2}\Pi _{3/2}$ state. 
To check whether the parasitic population of the 6p $^{2}\Pi _{3/2}$ state can explain the observed CsHe$^+$ yield in the measurements for the 6p $^{2}\Sigma _{1/2}$ state shown in Figure \ref{fig:allionTOFplots}f, we compared the effective overlap area between the absorption band and the laser excitation spectrum for two measurements. For the first measurement the excitation spectrum was centered at $E_{\mathrm{pump}}/(hc) = 11990\,$cm$^{-1}$, whereas for the second measurement the excitation spectrum was slightly red-detuned to increase the overlap with the high-energy tail of the 6p $^{2}\Pi _{3/2}$ dipole absorption band. Both the effective overlap and the CsHe$^+$ yield increased by the same amount for the red-detuned condition. Therefore, it is justified to attribute the observation of CsHe exciplexes at $E_{\mathrm{pump}}/(hc) = 11990$ cm$^{-1}$ to the parasitic population of the 6p $^{2}\Pi _{3/2}$ state.

For excitation via the 6p $^{2}\Sigma _{1/2}$ state, photoelectrons with a mean kinetic energy of 1.3\,eV are also observed (feature III in Figure \ref{fig:allelectronVMIplots}). At large pump-probe delays, this photoelectron yield is delay-independent. Compared to that, the photoelectron yield is increased by $\approx50\,\%$ at short pump-probe delays. The photoelectron yield from feature III is significantly increased when the excitation wavenumber is increased from $E_{\mathrm{pump}}/(hc) = 11900$\,cm$^{-1}$ to $E_{\mathrm{pump}}/(hc) = 11990$\,cm$^{-1}$. Unfortunately, we were not able to assign this measured photoelectron feature. Further investigations are necessary to disclose the underlying process which leads to the emission of photoelectrons at this specific kinetic energy. Because the overall signal intensity for feature III is small compared to the signal intensities for features I and II and the time constant for the decay of feature III (3 ps for excitation at $E_{\mathrm{pump}}/(hc) = 11900$\,cm$^{-1}$ and 7 ps for excitation at $E_{\mathrm{pump}}/(hc) = 11990$\,cm$^{-1}$) is quite different from the time constant obtained for the increase of feature I ($\tau^\mathrm{e}_{\mathrm{V,R}} =21$\,ps for excitation at $E_{\mathrm{pump}}/(hc) = 11900$\,cm$^{-1}$ and $\tau^\mathrm{e}_{\mathrm{V,R}}=23$\,ps for excitation at $E_{\mathrm{pump}}/(hc) = 11990$\,cm$^{-1}$), we believe that the contribution of this unknown reaction channel to the other dynamics is small.
\section{Summary}
Using pulsed femtosecond pump-probe spectroscopy in combination with electron-/ion-VMI and ion-TOF detection, we have resolved the picosecond desorption dynamics of excited Cs atoms and CsHe exciplexes attached to the surface of He nanodroplets. The dynamics were studied following the pulsed laser excitation of Cs-He$_N$ states close to the atomic 6p $\leftarrow$ 6s transitions in Cs and the subsequent detection of ions and/or electrons produced by a second, time-delayed ionization laser pulse.
The results of the ion-TOF measurements imply pump-probe-delay-dependent dynamics for both excited-state Cs atoms and CsHe exciplexes.

Our results indicate an impulsive-like desorption of Cs atoms after excitation of the 6p $^{2}\Pi _{1/2}$ state at an excitation wavenumber of $E_{\mathrm{pump}}/(hc)$ = 11260\,cm$^{-1}$. Compared to that, we have observed a significantly reduced ion yield and a high photoelectron yield at an excitation wavenumber of $E_{\mathrm{pump}}/(hc)$ = 11120\,cm$^{-1}$. The photoelectron yield resulting from the excitation at $E_{\mathrm{pump}}/(hc)$ = 11120\,cm$^{-1}$ is a factor of $\approx\,4.5$ higher than for excitation at $E_{\mathrm{pump}}/(hc)$ = 11260\,cm$^{-1}$. These findings confirm the results of previous experimental studies in which Cs atoms, excited at a wavenumber corresponding to the 6p $^{2}\Pi _{1/2} \leftarrow$ 6s $^{2}\Sigma _{1/2}$ transition in Cs-He$_N$, were found not to desorb from the He nanodroplet at excitation wavenumbers $E_{\mathrm{pump}}/(hc) < 11200$\,cm$^{-1}$. Following photoexcitation of the 6p $^{2}\Pi _{1/2}$ state and subsequent photo-ionization, we have measured only a very small number of CsHe$^+$ ions, corresponding to the desorption of CsHe exciplexes. This confirms the absence of exciplex formation in this state, due to a barrier which prevents access to the corresponding well in the CsHe potential\,\cite{Coppens.2019}. For the excitation of the 6p $^{2}\Pi _{1/2}$ state, the determined fall-back times are in qualitative agreement with the results from time-dependent $^4$He-DFT simulations by Coppens et al.\,\cite{Coppens.2019}. The experimentally obtained asymptotic ion velocity of the desorbing ions in the $^{2}\Pi _{1/2}$ state is higher than the theory value. This may be explained by signal contributions of the excitation at the blue edge of the dipole absorption spectrum, whereas the theory value is obtained at the maximum intensity of the dipole absorption spectrum. 

After excitation of the 6p $^{2}\Sigma _{1/2}$ state, we report an impulsive desorption of excited-state Cs atoms which is also supported by the results of the DFT simulations by Coppens et al.\,\cite{Coppens.2019}. Additionally, we have observed exceptionally high ion kinetic energies which cannot be explained by using the calculated pseudo-diatomic potential energy curve for this state. Our results strongly suggest that spin-orbit relaxation from the 6p $^{2}\Sigma _{1/2}$ state to the 6p $^{2}\Pi _{1/2}$ state leads to the desorption of the excited-state Cs atoms. We estimate that 36\,$\%$ of the energy released by spin-orbit relaxation is converted to the kinetic energy of the ions.

We have also observed the desorption of excited-state Cs atoms and CsHe exciplexes after excitation to the 6p $^{2}\Pi _{3/2}$ state, which is not predicted by theory. 
Because the observed pump-probe dependence for the excitation of this state is very different compared to the other states, we believe that a mechanism is at play which is distinctively different from impulsive desorption. For this state, we suggest an evaporative-like desorption mechanism.
Furthermore, we show that the vibrational relaxation of CsHe exciplexes in the 6p $^{2}\Pi _{3/2}$ state can release enough energy to cause the desorption of CsHe exciplexes from the droplet surface. The same process was proposed before for RbHe exciplexes\,\cite{Leino.2011} and NaHe exciplexes\,\cite{Loginov.2008}.
CsHe exciplexes formed after the excitation of the 6p $^{2}\Sigma _{1/2}$ state show impulsive dynamics which are also not predicted by theory. We explain this observation by a parasitic population of the 6p $^{2}\Pi _{3/2}$ state as a result of the broad laser excitation spectrum.

The desorption dynamics of Cs$_2$, which are observed as an increase of the Cs$_2^+$ ion yield as a function of pump-probe delay, are in the same time range as those observed for excited-state Cs atoms and CsHe exciplexes. Additional theoretical studies are necessary to understand the dynamics observed for Cs$_2$ as well as their possible role in the CsHe exciplex dynamics observed after excitation of the 6p $^{2}\Sigma _{1/2}$ state. 
\section{Acknowledgements}
This work was financially supported by the German Research Foundation (DFG; GRK 2079 `Cold Controlled Ensembles in Physics and Chemistry') and by the European Research Council (ERC) Advanced Grant COCONIS (694965). K.D.\ acknowledges financial support by the Fonds der Chemischen Industrie (Liebig Fellowship).
The authors declare no competing financial interest.
\bibliography{Cs-dynamic-manuscript.bib}
%\printbibliography
%

%
\end{document}